\documentclass[12pt,letter]{article}
\pdfoutput=1
\usepackage{graphicx, epsfig, color,cite}
\usepackage{amsmath,amsthm,amsfonts,geometry,etoolbox}

\usepackage{amssymb}
\usepackage{caption,subcaption,graphicx,appendix}
\usepackage[hidelinks]{hyperref}
\def\to{\rightarrow}

\def\bi{\begin{itemize}}
\def\ei{\end{itemize}}

\def\tst{\tilde t}

\def\tg{\tilde g}

\def\tell{\tilde\ell}
\def\tq{\tilde q}
\def\alt{\lesssim}
\def\agt{\gtrsim}
\def\be{\begin{equation}}  
\def\ee{\end{equation}}  
\def\bea{\begin{eqnarray}}  
\def\eea{\end{eqnarray}}

\newcommand{\myeq}{\begin{small}\begin{equation}\begin{aligned}}
\newcommand{\myeqend}{\end{aligned}\end{equation}\end{small}}

\begin{document}
\begin{titlepage}
\begin{flushright}
OU-HEP-240301
\end{flushright}

\vspace{0.5cm}
\begin{center}
  {\Large \bf Weak scale supersymmetry\\
    emergent from the string landscape}\\

\vspace{1.2cm} \renewcommand{\thefootnote}{\fnsymbol{footnote}}
{\large Howard Baer$^{1}$\footnote[1]{Email: baer@ou.edu },
Vernon Barger$^2$\footnote[2]{Email: barger@pheno.wisc.edu},
Dakotah Martinez$^1$\footnote[4]{Email: dakotah.s.martinez-1@ou.edu} and
Shadman Salam$^3$\footnote[4]{Email: ext.shadman.salam@bracu.ac.bd} 
}\\ 
\vspace{1.2cm} \renewcommand{\thefootnote}{\arabic{footnote}}
{\it 
$^1$Homer L. Dodge Department of Physics and Astronomy,\\
University of Oklahoma, Norman, OK 73019, USA \\[3pt]
}
{\it 
$^2$Department of Physics,
University of Wisconsin, Madison, WI 53706 USA \\[3pt]
}
{\it 
$^3$Department of Mathematics and Natural Sciences,
Brac University, Dhaka 1212, Bangladesh \\[3pt]
}

\end{center}

\vspace{0.5cm}
\begin{abstract}
\noindent
Superstring flux compactifications can stabilize all moduli while
leading to an enormous number of vacua solutions,
each leading to different $4-d$ laws of physics.
While the string landscape provides at present the only plausible explanation
for the size of the cosmological constant, it may also predict the form of
weak scale supersymmetry which is expected to emerge.
Rather general arguments suggest a power-law draw to large soft terms,
but these are subject to an anthropic selection of not-too-large a
value for the weak scale. The combined selection allows one to compute relative
probabilities for the emergence of supersymmetric models from the landscape.
Models with weak scale naturalness appear most likely to emerge since
they have the largest parameter space on the landscape.
For finetuned models such as high scale SUSY or split SUSY,
the required weak scale finetuning shrinks their parameter space to
tiny volumes, making them much less likely to appear compared to natural models.
Probability distributions for sparticle and Higgs masses from natural models
show a preference for Higgs mass $m_h\sim 125$ GeV with sparticles typically
beyond present LHC limits, in accord with data.
From these considerations, we briefly describe how natural SUSY is expected
to be revealed at future LHC upgrades.
This article is a contribution to the Special Edition of the journal
{\it Entropy} honoring Paul Frampton on his 80th birthday.

\end{abstract}
\end{titlepage}

\section{Introduction}
\label{sec:intro}

Superstring theory provides the most promising avenue for unifying
the Standard Model with gravity, but at the cost of requiring
6 or 7 extra spatial dimensions\cite{Green:1987sp,Green:1987mn,Polchinski:1998rq,Polchinski:1998rr,Hebecker:2021egx}.
The low energy limit $E<m_P$ (where $m_P$ is the reduced Planck mass)
of string theory, once Kaluza-Klein modes are
integrated out, is expected to be $10-d$ supergravity (SUGRA).
The $10-d$ SUGRA theory is then assumed to be compactified to a tiny $6-d$
space $K$ tensored with our usual $4-d$ (approximately) Minkowski
spacetime $M_4$: $M_{10}=M_4\times K$.
Originally, $K$ was taken to be a $6-d$ compact Ricci-flat
K\"ahler manifold with special holonomy\cite{Candelas:1985en};
such a Calabi-Yau manifold admits a conserved
Killing spinor leading to a remnant $N=1$ supersymmetry (SUSY) on $M_4$.

The cosmological constant (CC) problem remained a thorny issue until the
early 2000s when it was realized that string flux compactifications
could lead to an enormous number of vacuum states each with different
$4-d$ laws of physics, and in particular, different $\Lambda_{CC}$
values\cite{Bousso:2000xa}.
Such large numbers of vacuum states ($N_{vac}\sim 10^{500}$ is an oft-quoted
number\cite{Ashok:2003gk}) provided a setting for Weinberg's
anthropic solution to the CC problem\cite{Weinberg:1987dv}.
But if the landscape\cite{Susskind:2003kw} of string vacua provides
a solution to the CC problem, might it also enter into other naturalness
problems, such as the $m_{weak}/m_P$ (or related, $m_{SUSY}/m_P$)
hierarchy problems (where $m_P\simeq 2.4\times 10^{18}$ GeV)? 

In this contribution to the  volume of the journal
{\it Entropy} honoring Paul Frampton on his 80th birthday,
we address this question.
Here, we will put forward
arguments in favor not only of weak scale SUSY as emergent from the
string landscape, but indeed of a special form of weak scale SUSY (WSS)--
SUSY with radiatively-driven naturalness\cite{Baer:2012up,Baer:2012cf},
or stringy-natural SUSY\cite{Baer:2019cae}.
The specific form of WSS predicts at present that a light SUSY
Higgs boson with mass $m_h\simeq 125$ GeV should emerge, whilst
sparticles masses are at present somewhat or well beyond reach of the
CERN Large Hadron Collider (LHC)\cite{Baer:2017uvn}.
It also allows us to predict a variety of SUSY signatures which
may allow for SUSY discovery at
LHC luminosity upgrades over the upcoming years.
Perhaps most important of these are the soft isolated opposite-sign
dileptons+MET which arise from light higgsino pair production\cite{Baer:2011ec} and
which recoil against a hard initial state jet
radiation\cite{Han:2014kaa,Baer:2014kya,Han:2015lma,Baer:2021srt}.
At present, both ATLAS\cite{ATLAS:2019lng} and CMS\cite{CMS:2021edw}
with 139 fb$^{-1}$
seem to have $2\sigma$ excesses in this channel, and an associated
monojet signal may also be emerging\cite{Agin:2023yoq}.

\section{Approximate supersymmetric vacua from string theory}

The main motivation for SUSY is that it provides a 'tHooft
technical naturalness solution to the gauge hierarchy problem
via the cancellation of quadratic divergences associated with the Higgs sector.
This is true for SUSY breaking at any energy scale below $m_P$, since
in the limit of $m_{SUSY}\to 0$, the model becomes more (super)symmetric.
Thus, SUSY provides a technically natural solution to the so-called
Big Hierarchy problem.

Specific motivation for weak scale SUSY
comes from the Little Hierarchy problem and what we call
{\it practical naturalness\cite{Baer:2015rja,Baer:2023cvi}:} an observable ${\cal O}$ is practically
natural if all independent contributions to ${\cal O}$ are comparable to
or less than ${\cal O}$. For the case of WSS, we can relate the weak scale
$m_{weak}\sim m_{W,Z,h}\sim 100$ GeV to the weak scale soft SUSY breaking terms
and SUSY conserving $\mu$ term via the scalar potential minimization
conditions:
\be
\frac{m_Z^2}{2}=\frac{m_{H_d}^2+\Sigma_d^d-(m_{H_u}^2+\Sigma_u^u)\tan^2\beta}{\tan^2\beta -1}-\mu^2\sim -m_{H_u}^2-\Sigma_u^u(\tst_{1,2})-\mu^2
\label{eq:mzs}
\ee
where $m_{H_{u,d}}^2$ are the soft SUSY breaking Higgs masses and the
$\Sigma_{u,d}^{u,d}$ contain an assortment of loop corrections to the scalar
potential (explicit formulae are included in Ref's \cite{Baer:2012cf}).
A measure of practical naturalness $\Delta_{EW}$ can be defined which compares
the largest (absolute) contribution to the right-hand-side of Eq. \ref{eq:mzs}
to $m_Z^2/2$. Requiring $\Delta_{EW}\alt 30$ fulfills the practical naturalness
condition.
From Eq. \ref{eq:mzs}, we see immediately that $m_{H_u}^2$ must be
driven to {\it small} negative values at the weak scale while the
$\mu$ term must also be $\mu \sim 100-350$ GeV. The latter condition means
the higgsinos are usually the lightest SUSY particles, and the only ones
required to be $\sim m_{weak}$. The other sparticles enter via the
$\Sigma_{u,d}^{u,d}$ terms and hence are suppressed by loop factors, and so can
live in the TeV or beyond range.
We shall see shortly that practical naturalness is closely linked to
selection of SUSY models on the landscape.

On the theory side, we expect the $4-d$ vacua emergent from the landscape
to often contain some remnant SUSY.
\bi
\item {\it Remnant spacetime SUSY:}
  In Ref. \cite{Acharya:2019mcu}, Acharya argues that all stable, Ricci-flat
  manifolds in dimensions $<11$ have special holonomy, and consequently a
  conserved Killing spinor. If so, then
  some remnant spacetime SUSY should exist in the $4-d$ low-energy effective field theory (LE-EFT).
\item {\it EW stability:} A problem for the Standard Model to be the
  low-energy effective field theory for $E<m_P$ is electroweak stability in
  that the Higgs quartic term $\lambda$ may evolve to negative values at
  $E>m_{GUT}$ leading to a runaway scalar potential.
  For $m_t\sim 173.2$ GeV, the SM is just on the edge of
  metastability/runaway\cite{Degrassi:2012ry,Buttazzo:2013uya}.
  The Minimal Supersymmetric Standard Model (MSSM) has no such problem
  since for the MSSM the quartic couplings involve the gauge couplings
  which are always positive.
\item {\it Landscape vacua stability:} In Ref's \cite{Dine:2007er,Dine:2009tv},
  Dine {\it et al.} ask
  what sort of conditions can stabilize landscape deSitter vacua against
  decay to AdS vacua, leading to a big crunch.
  The presence of SUSY leads to absolutely stable vacua whilst the presence of
  approximate (broken) SUSY leads to (metastable) vacua decay rates
  $\Gamma\sim m_P e^{-m_P^2/m_{3/2}^2}$ far beyond the age of the universe.
\item {\it Hierarchy of scales:} While a hierarchy of scales is typically hard to come by in many
  BSM models, SUSY models allow for dynamical SUSY breaking\cite{Dine:2010cv}
  where the SUSY breaking scale $m_{hidden}$ is gained via
  dimensional transmutation $m_{hidden}\sim m_P e^{-8\pi^2/bg^2}$ and
  where the soft terms are developed as $m_{soft}\sim m_{hidden}^2/m_P$
  under gravity-mediation.\footnote{Here, we only consider gravity-mediation
    since gauge mediation leads to tiny trilinear soft  terms $A$ which
    then require unnatural top-squark contributions $\Sigma_u^u$
    to gain $m_h\sim 125$ GeV\cite{Baer:2014ica}.}
\item {\it Harmony:} Witten emphasizes that consistent QFTs exist for
  spin-0, 1/2, 1, 3/2 and 2. The graviton is the physical spin-2 particle
  and the spin-3/2 Rarita-Schwinger gravitino field would exist as
  the superpartner of the graviton, thus filling out all allowed spin states.
  \ei

In addition, WSS is motivated experimentally by a variety of measurements.
\bi
\item The measured values of the gauge couplings unify under MSSM
  RG evolution while not under most other BSM extensions,
  including the SM itself\cite{Dimopoulos:1981yj}.
\item The measured top-quark mass is large enough to seed the required
  radiative breakdown of EW symmetry\cite{Ibanez:1982fr}.
\item The measured value of $m_h\simeq 125$ GeV falls squarely into the
  range allowed by the MSSM: $m_h\alt 130$ GeV\cite{Slavich:2020zjv}.
\item Precision EW corrections tend to prefer the (heavy spectra) MSSM
  over the SM\cite{Heinemeyer:2006px}.
\ei
It is often complained, with good reason, that gravity mediation has its own
flavor and CP problems, the former arising from operators such as
$\int d^4\theta S^\dagger S Q_i^{\dagger}Q_j/m_P^2$ where $S$ is a hidden sector
superfield obtaining a SUSY breaking vev $F_S\sim 10^{11}$ GeV and the
$Q_i$ are visible sector chiral superfields with generation index $i,j=1-3$.
Since no symmetry forbids such flavor mixing, then FCNCs
are expected to be large in gravity-mediated SUSY breaking (historically,
this strongly motivated the search for flavor conserving models such as
gauge-mediation and sometimes anomaly-mediation.) It is pointed out
in Ref. \cite{Baer:2019zfl} that the landscape provides its own
decoupling/quasi-degeneracy solution to the SUSY flavor and CP problems
by pulling first/second generation matter scalars to a flavor-independent upper
bound in the 20-40 TeV range.

For these reasons, we will assume a so-called ``fertile patch''
or friendly neighborhood\cite{Arkani-Hamed:2005zuc} of the string landscape:
those vacua which include the MSSM as the LE-EFT and where only the CC
and the magnitude of the weak scale scan within the landscape.
In this case, Yukawa couplings and gauge couplings are instead fixed by
string dynamics rather than environmental selection. This leads to
predictive landscape models\cite{Arkani-Hamed:2005zuc}:
if the CC is too large, then large scale
structure will not form which seems required for complexity to emerge
(the structure principle, leading to Weinberg's successful prediction
of $\Lambda_{CC}$). Only the magnitude of the weak scale scans.
If $m_{weak}^{PU}\agt 4m_{weak}^{OU}$, then the down-up quark mass difference
becomes so large that neutrons are no longer stable in nuclei and the only
atoms formed in the early universe are Hydrogen.
If $m_{weak}^{PU}\alt  0.5m_{weak}^{OU}$, then we get a universe with only neutrons.
This is the atomic principle\cite{Agrawal:1998xa},
since complex nuclei are also apparently needed for complexity to emerge
in any  pocket universe (PU) within the greater multiverse (and
where OU refers to $m_{weak}$ in our universe).

\section{Natural SUSY from the landscape}

It is emphasized by Douglas that the CC scans independently of the SUSY
breaking scale in the landscape\cite{Douglas:2004qg}.
For the SUSY breaking scale, we expect the vacua to be distributed as
\be
dN_{vac}\sim f_{SUSY}\cdot f_{EWSB}\cdot dm_{SUSY}^2
\label{eq:dNvac}
\ee
where $m_{SUSY}$ is the overall hidden sector SUSY breaking scale
expected to be $\sim 10^{11}$ GeV such that the scale of soft terms is
given by $m_{soft}\sim m_{SUSY}^2/m_P$.

\subsection{Distribution of soft breaking terms on the landscape}

How is $f_{SUSY}$ distributed? Douglas\cite{Douglas:2004qg} emphasizes
that there is nothing in string theory to favor any particular
SUSY breaking vev over another,
and hence $m_{soft}$ would be distributed as a power-law
\be
f_{SUSY}\sim m_{soft}^{2n_F+n_D-1}
\label{eq:fsusy}
\ee
where $n_F$ is the total number of (complex-valued) $F$-breaking fields
and $n_D$ is the total number of (real-valued) $D$-breaking fields
contributing to the overall scale of SUSY breaking
$m_{SUSY}^4=\sum_i |F_i|^2+\sum_\alpha D_\alpha^2$.
The prefactor of 2 in the exponent comes since the $F_i$ are distributed
randomly as complex numbers. 
For the textbook case of
SUSY breaking via a single $F$ term, then we expect $f_{SUSY}\sim m_{soft}^1$,
{\it i.e.} a {\it linear} draw to large soft terms.
If more hidden fields contribute to the overall SUSY breaking scale, then the
draw to large soft terms will be a stronger power-law.

While the overall SUSY breaking scale is distributed as a power-law,
the different functional dependence\cite{Soni:1983rm,Kaplunovsky:1993rd,Brignole:1993dj} of the soft terms on the
hidden sector SUSY breaking fields means that gaugino masses,
the trilinear soft terms and the various scalar masses will effectively scan
independently on the landscape\cite{Baer:2020vad}. Now it is an {\it advantage}
that different scalar mass-squared terms scan independently
(as expected in SUGRA) since the first/second generation scalars get pulled
to much higher values than 3rd generation scalars, while the two Higgs
soft masses are also non-universal and scan independently.
This situation is borne out in Nilles {\it et al.} mini-landscape
where different fields gain different soft masses due to their different
geographical locations on the compactification manifold\cite{Nilles:2014owa}.
In terms of gravity mediation, then the so-called $n$-extra-parameter
non-universal Higgs model (NUHMn) with parameters\cite{Ellis:2002iu,Baer:2005bu}
\be
m_0(i),\ m_{H_u},\ m_{H_d},\ m_{1/2}, A_0,\ \tan\beta\ \ \ \ \ (NUHM4)
\label{eq:nuhm4}
\ee
provides the proper template. Since the matter scalars fill out a complete
spinor rep of $SO(10)$, we assume each generation $i=1-3$
is unified to $m_0(i)$. Also, for convenience one may ultimately trade
$m_{H_u}$ and $m_{H_d}$ for the more convenient weak scale parameters
$m_A$ and $\mu$. One may also build (and scan separately) the natural
anomaly-mediated SUSY breaking model\cite{Baer:2018hwa,Baer:2023ech} (nAMSB) and the
natural generalized mirage mediation model\cite{Baer:2016hfa} (nGMM).

\subsection{The ABDS window}

The anthropic selection on the landscape comes from $f_{EWSB}$.
This involves a rather unheralded prediction of the MSSM: the value of the
weak scale in terms of soft SUSY breaking parameters and $\mu$,
as displayed in Eq. \ref{eq:mzs}. However, in the multiverse, here we rely
on the existence of a friendly neighborhood\cite{Arkani-Hamed:2005zuc}
wherein the LE-EFT contains the MSSM but where only dimensionful
quantities such as $\Lambda_{CC}$
and $v_u^2+v_d^2$ scan, whilst dimensionless quantities like gauge and
Yukawa couplings are determined by dynamics. This assumption leads to
{\it predictivity} as we shall see.

Under these assumptions, then we ask what conditions lead to complex nuclei,
atoms as we know them, and hence the ability to generate complex lifeforms
in a pocket universe? For different values of soft terms, frequently one
is pushed into a weak scale scalar potential with charge-or-color
breaking minima (CCB) where one or more charged or colored scalar mass squared
is driven tachyonic (i.e., $m^2<0$). Such CCB minima must be vetoed.
Also, for too large of values of $m_{H_u}^2$, then its value is {\it not}
driven to negative values and EW symmetry is not broken. These we label as
``no EWSB" and veto them as well. In practice, we must check boundedness of the scalar potential from below in the vacuum stability conditions and that the origin of field space has been destabilized at tree-level. 

\begin{figure}[t]
  \centering
  {\includegraphics[width=0.7\textwidth]{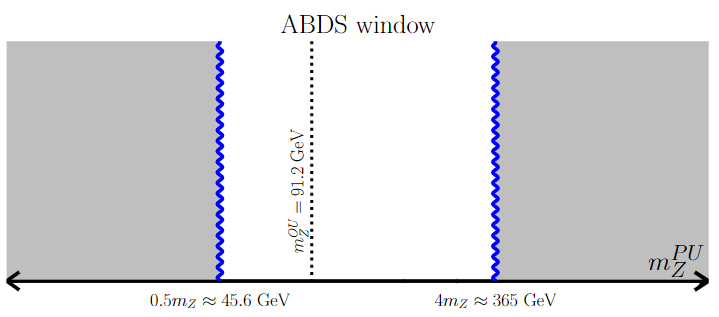}}\quad
  \caption{The ABDS-allowed window within the range of $m_Z^{PU}$ values.}  
\label{fig:ABDS}
\end{figure}

At this point, we are left with (MS)SM vacua
where EW symmetry is properly broken, but where $m_{weak}\sim m_{W,Z,h}$ is
at a different value from what we see in our universe.
Here, we rely on the prescient analysis of Agrawal, Barr, Donoghue and
Seckel (ABDS)\cite{Agrawal:1997gf,Agrawal:1998xa}.
If the derived value of the weak scale is bigger than ours
by a factor $(2-5)$, then the light quark mass difference $m_d-m_u$
becomes so large that neutrons are no longer stable in the nucleus
and nuclei with $Z\gg N$ are not bound; such  pocket universes
would have nuclei of single protons only, and would be chemically inert.
Likewise, if the PU value of the weak scale is a factor $\sim 0.5$ less than
our measured value, then one obtains a universe with only neutrons-- also
chemically inert. The ABDS window of allowed values is that
\be
0.5 m_{weak}^{OU}<m_{weak}^{PU} \lesssim 4 m_{weak}^{OU}
\label{eq:abds}
\ee
where we take the $(2-5)m_{weak}^{OU}$ to be $\sim 4m_{weak}^{OU}$ for
definiteness, which is probably a conservative value.
It is very central to our analysis and so is displayed in Fig. \ref{fig:ABDS}.
Our anthropic condition $f_{EWSB}$ is then that the scalar potential
lead to minima with not only appropriate EWSB, but also that the derived
value of the weak scale lie within the ABDS window.
Vacua not fulfilling these conditions must be vetoed.\footnote{Early papers
  on this topic used instead a naturalness ``penalty'' of
  $f_{EWSB}\sim m_{weak}^2/m_{SUSY}^2$; this condition would allow for
  many of the vacua which are forbidden by our approach.}

\subsection{EW natural SUSY emergent from the landscape}

The next goal is to build a toy simulation of our friendly neighborhood of
the string landscape. We can generate the soft terms of Eq. \ref{eq:nuhm4}
according to a power-law selection, usually taken to be $n=2n_F+n_D-1=1$
(linear draw). While Eq. \ref{eq:fsusy} favors the largest possible soft terms,
the anthropic veto $f_{EWSB}$ places an upper bound on such terms because
usually large soft terms lead to too large a value of $m_{weak}^{PU}$
beyond the ABDS window. The trick is to take the upper bound on scan limits
beyond the upper bound posed by $f_{EWSB}$. However, in some cases larger soft
terms are {\it more} apt to lie within the ABDS window. A case in point is
$m_{H_u}^2$: the smaller its value, the deeper negative it runs to unnatural
values at the weak scale, while as it gets larger, then it barely runs
negative: EW symmetry is barely broken. As its high scale value becomes
even larger, it doesn't run negative by $m_{weak}$, and EW symmetry is not
broken-- such vacua are vetoed. Also, for small $A_0$,
the $\Sigma_u^u(\tst_{1,2})$ terms can be large. When $A_0$ becomes large
negative, then canellations occur in $\Sigma_u^u(\tst_{1,2})$ such that
these loop corrections then lie within the ABDS window: large negative
weak scale $A$ terms make $\Sigma_u^u(\tst_{1,2})$ more natural while
raising the light Higgs mass $m_h\sim 125$ GeV.

\begin{figure}[h!]
\centering
   \includegraphics[width=0.8\textwidth]{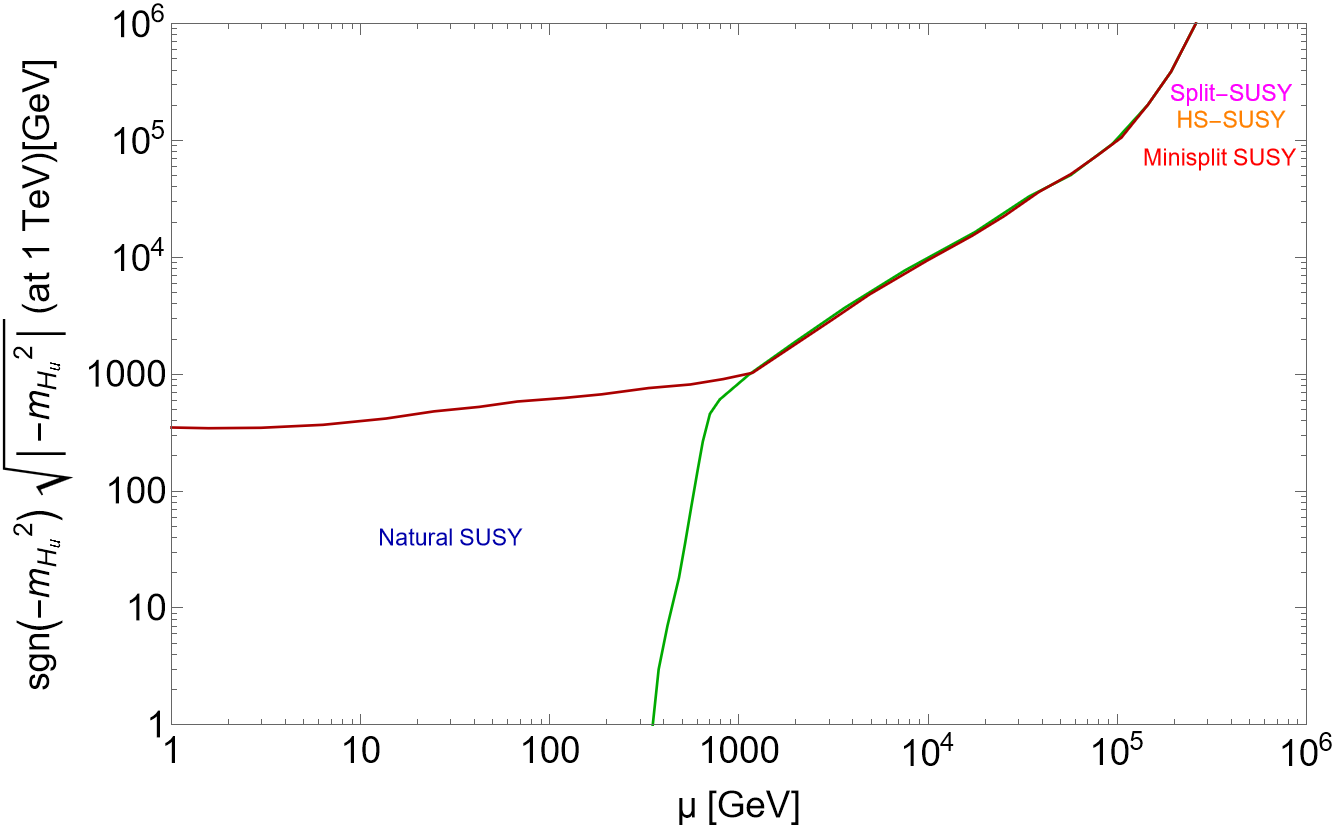}\quad
   \caption{The $\mu^{PU}$ vs. $\sqrt{-m_{H_u}^2(weak)}$ parameter
space in a toy model ignoring radiative corrections to the
Higgs potential. The region between red and green curves
leads to $m_{weak}^{PU}<4 m_{weak}^{OU}$ so that the atomic principle 
is satisfied.
}
    \label{fig:mu_mhu}   
\end{figure}

A plot of the weak scale values of $m_{H_u}$ and $\mu$ is shown in
Fig. \ref{fig:mu_mhu} (taken from Ref. \cite{Baer:2022wxe})
for the case where all radiative corrections--
some negative and some positive\cite{Baer:2021tta}-- lie within the ABDS window.
The ABDS window lies between the red and green curves.
Imagine playing darts with this target, trying to land your dart within
the ABDS window. There is a large region to the lower-left where both
$m_{H_u}$ and $\mu$ are $\alt 350$ GeV which leads to PUs with complexity.
Alternatively, if you want to land your dart at a point with $\mu\sim 1000$ GeV,
then the target space has pinched off to a tiny volume: the target space
is finetuned and your dart will almost never land there. The EW natural
SUSY models live in the lower-left ABDS window while finetuned SUSY models
with large $\Delta_{EW}$ lie within the extremely small volume between the red and green curves in the upper-right plane.
This tightly-constrained region is labeled by split
SUSY\cite{Arkani-Hamed:2004ymt}, high scale SUSY\cite{Barger:2005qy}
and minisplit\cite{Arvanitaki:2012ps}.

It is often said that landscape selection offers an alternative to naturalness
and allows for finetuned SUSY models. After all, isn't the CC finetuned?
However, from Fig. \ref{fig:mu_mhu} we see that models with EW naturalness
(low $\Delta_{EW}$) have a far greater relative probability to emerge from
landscape selection than do finetuned SUSY models.

\begin{figure}
    \centering
    \includegraphics[width=0.8\textwidth]{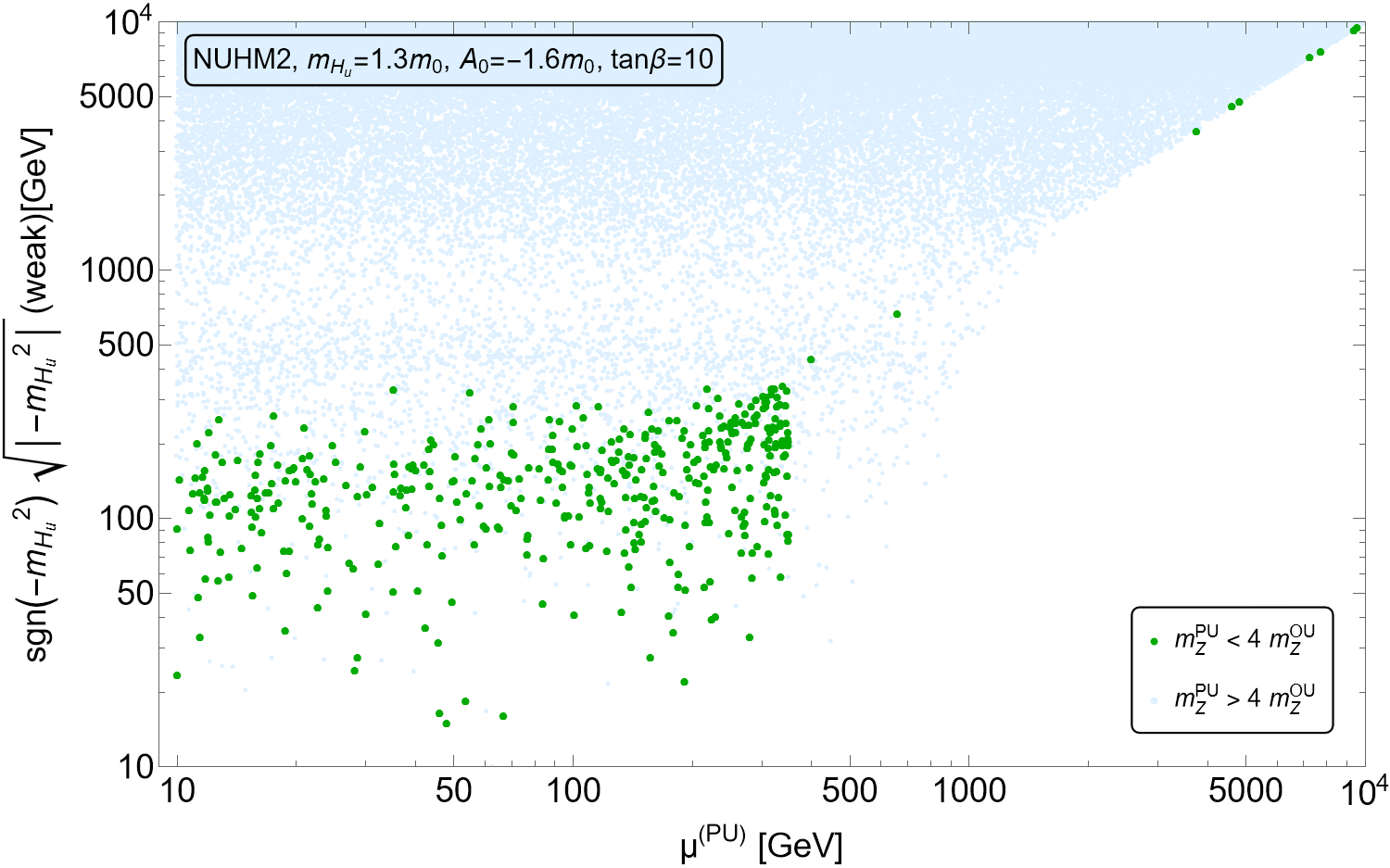}\quad
    \captionof{figure}{The value of $m_{H_u}(weak)$ vs. $\mu^{PU}$
The green points denote vacua with appropriate
EWSB and with $m_{weak}^{PU}<4 m_{weak}^{OU}$ so that the atomic principle 
is satisfied. Blue points have $m_{weak}^{PU}>4 m_{weak}^{OU}$.
}
\label{fig:dots1}
\end{figure}
   
In Fig. \ref{fig:dots1} (from Ref. \cite{Baer:2022wxe}),
we perform a numerical exercise to generate high scale
SUSY soft terms in accord with an $n=1$ draw in Eq. \ref{eq:fsusy}. 
The green dots are viable vacua states with appropriate EWSB and
$m_{weak}^{PU}$ within the ABDS window.
While some dots do land in the finetuned region, the bulk of points
lie within the EW natural SUSY parameter space.

\begin{figure}[!htbp]
\begin{center}
\includegraphics[height=0.4\textheight]{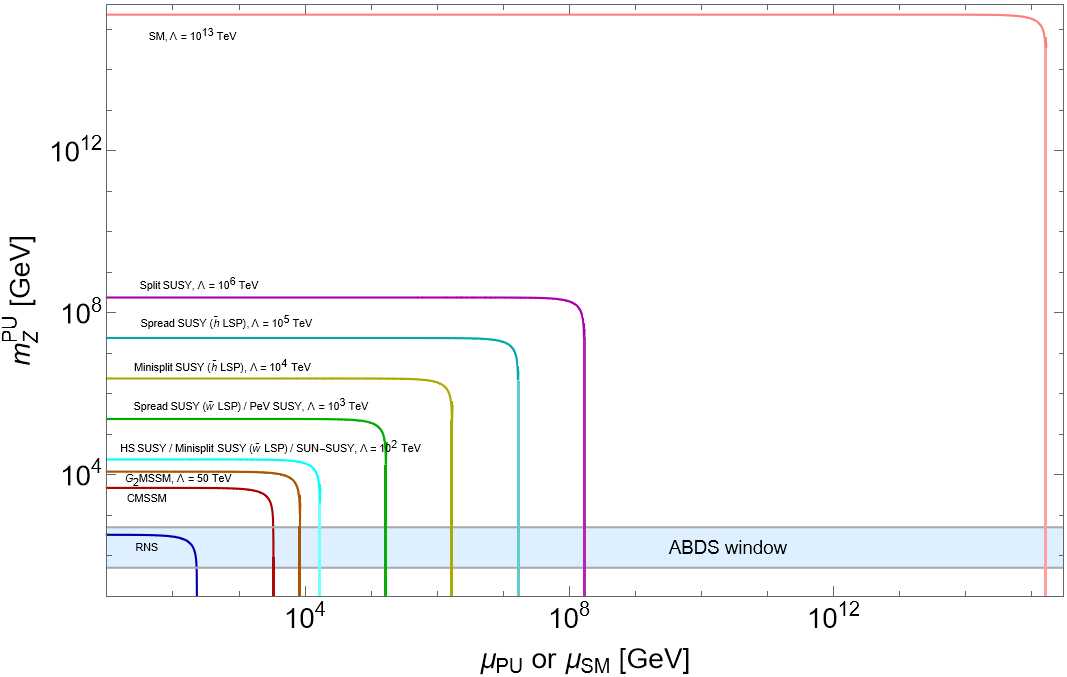}
\caption{Values of $m_Z^{PU}$ vs. $\mu_{PU}$ or $\mu_{SM}$ for
  various natural (RNS) and unnatural SUSY models and the SM.
  The ABDS window extends here from $m_Z^{PU}\sim 50-500$ GeV.
\label{fig:mzPU}}
\end{center}
\end{figure}

An alternative view is gained from Fig. \ref{fig:mzPU} from
Ref. \cite{Baer:2022dfc}. Here, we compute contributions to the scalar potential within a variety of SUSY models
including RNS (radiatively-driven natural SUSY\cite{Baer:2012up}),
CMSSM\cite{Kane:1993td}, G$_2$MSSM\cite{Acharya:2008zi},
high scale SUSY\cite{Barger:2007qb}, spread SUSY\cite{Hall:2011jd},
minisplit\cite{Arvanitaki:2012ps}, split SUSY\cite{Arkani-Hamed:2004ymt}
and the SM with cutoff $\Lambda =10^{13}$ TeV,
indicative of the neutrino see-saw scale\cite{Vissani:1997ys}.
The $x$-axis is either the SM $\mu$ parameter or the SUSY $\mu$ parameter
while the $y$-axis is the calculated value of $m_Z$ within the PU.
The ABDS window is the horizontal blue-shaded region.
For $\mu$ distributed as equally likely at all scales (integrates to a log),
then the length of the $x$-axis interval leading to $m_Z^{PU}$ within
the ABDS window can be regarded as a relative probability measure $P_\mu$ for
the model to emerge from the landscape. There is a substantial interval for
the RNS model, but for finetuned SUSY models, the interval is typically
much more narrow than the width of the printed curves.
We can see that finetuned models have only a tiny range of $\mu$ values
which allow habitation within the ABDS window.
Using the magic of algebra, the width of the $\mu$ intervals can be computed,
and the results are in Table \ref{tab:models}. Here, $P_\mu$ is to be considered as a {\it relative} probability. From the Table, we see that the SM is
about $10^{-27}$ times likely to emerge as compared to RNS.
Minisplit is $10^{-4}- 10^{-8}$ times likely to emerge (depending on the
version of minisplit).
Even the once-popular CMSSM model is $\sim 10^{-3}$ times likely than RNS to
emerge from the landscape.
\begin{table}\centering
\begin{tabular}{lcccccc}
\hline
model & $\tilde{m}(1,2)$ & $\tilde{m}(3)$ & gauginos & higgsinos & $m_h$ & $P_\mu$ \\
\hline
SM & - & - & - & -& - & $7\cdot10^{-27}$ \\
CMSSM ($\Delta_{EW}=2641$) & $\sim 1$ & $\sim 1$ & $\sim 1$ & $\sim 1$ & $0.1-0.13$
& $5\cdot 10^{-3}$ \\
PeV SUSY & $\sim 10^3$ & $\sim 10^3$ & $\sim 1$ & $1-10^3$ &
$0.125-0.155$ & $5\cdot 10^{-6}$ \\
Split SUSY & $\sim 10^6$ & $\sim 10^6$ & $\sim 1$ & $\sim 1$ & $0.13-0.155$
& $7\cdot 10^{-12}$ \\
HS-SUSY & $\agt 10^2$ & $\agt 10^2$ & $\agt 10^2$ & $\agt 10^2$ & $0.125-0.16$
& $6\cdot 10^{-4}$ \\
Spread ($\tilde{h}$LSP) & $10^{5}$  & $10^5$ & $10^2$ & $\sim 1$ & $0.125-0.15$ & $9\cdot 10^{-10}$ \\
Spread ($\tilde{w}$LSP) & $10^{3}$ & $10^{3}$ & $\sim 1$ & $\sim 10^2$ & $0.125-0.14$  & $5\cdot 10^{-6}$ \\
Mini-Split ($\tilde{h}$LSP)& $\sim 10^4$ & $\sim 10^4$ & $\sim 10^2$ & $\sim 1$  & $0.125-0.14$ & $8\cdot10^{-8}$ \\
Mini-Split ($\tilde{w}$LSP)& $\sim 10^2$ & $\sim 10^2$ & $\sim 1$ & $\sim 10^2$ & $0.11-0.13$ & $4\cdot 10^{-4}$ \\
SUN-SUSY  & $\sim 10^2$ & $\sim 10^2$ & $\sim 1$ & $\sim 10^2$  & $0.125$
& $4\cdot 10^{-4}$ \\
G$_2$MSSM  & $30-100$ & $30-100$ & $\sim 1$  & $\sim 1$  & $0.11-0.13$
& $2\cdot 10^{-3}$ \\
RNS/landscape & $5-40$  & $0.5-3$ & $\sim 1$ & $0.1-0.35$ & $0.123-0.126$
& $1.4$ \\
\hline
\end{tabular}
\caption{A survey of some unnatural and natural SUSY models
  along with general expectations for sparticle and Higgs
  mass spectra in TeV units.
  We also show relative probability measure $P_\mu$ for the model to emerge
  from the landscape.
  For RNS, we take $\mu_{min}=10$ GeV.
}
\label{tab:models}
\end{table}

\section{Radiatively-driven natural SUSY}
\label{sec:rns}

Along with probability distributions for models to emerge from the landscape,
one can compute probability distributions for sparticle and Higgs mass
values from particular models given an assumed value of $n$ in $f_{SUSY}$.
Here, we use a linear draw, $n=1$, to large soft terms with the NUHM4 model
as the LE-EFT. We capture non-finetuned models by requiring
$\Delta_{EW}\alt 30$, {\it i.e.}, that the largest independent contribution
to $m_Z$ lies within the ABDS window.
These models have radiatively-driven naturalness (RNS) where RG running
drives various soft terms to natural values at the weak scale.

\begin{figure}[th!]
\begin{center}
\includegraphics[height=0.3\textheight]{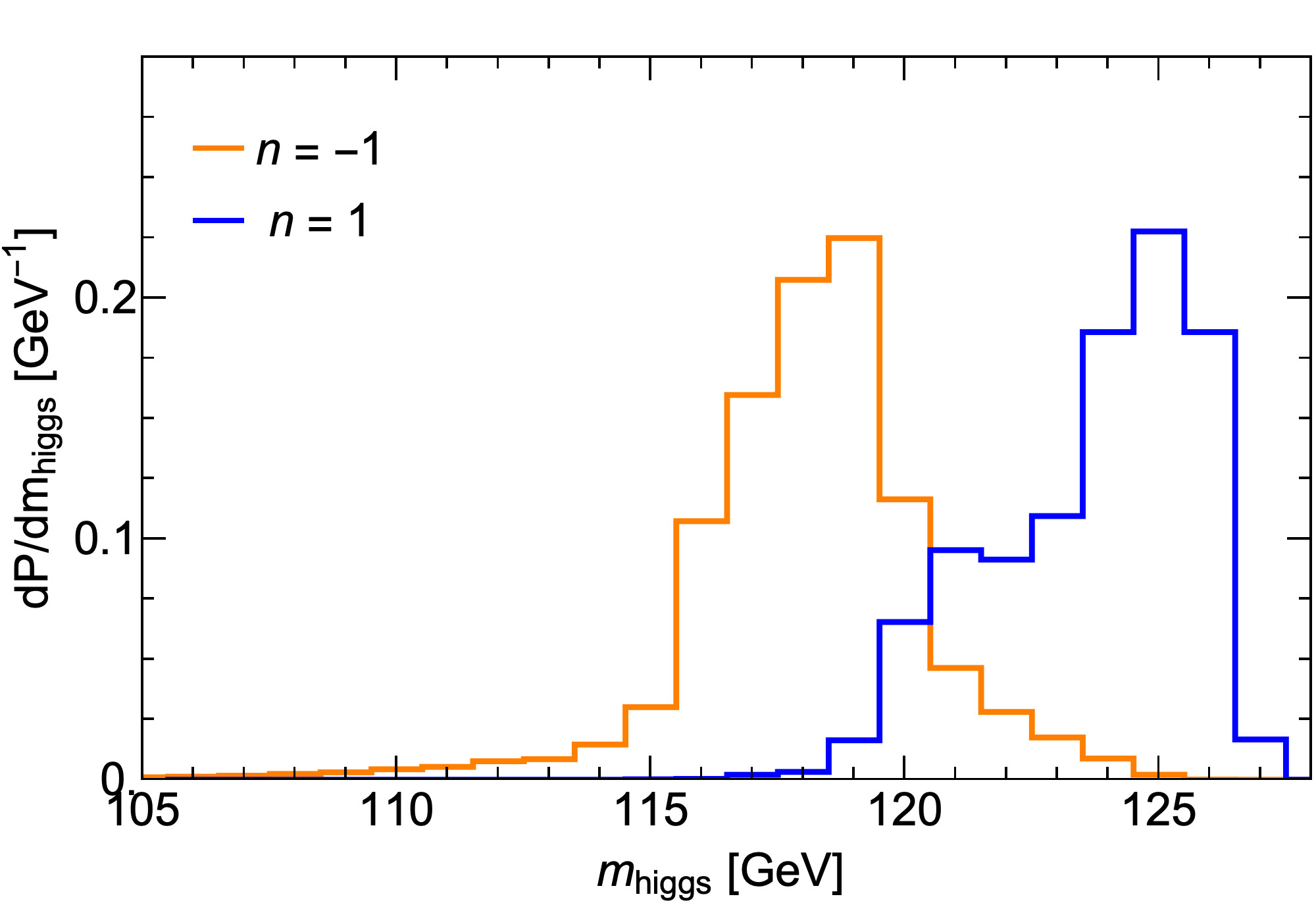}
\caption{Probability distributions for the light Higgs scalar mass $m_h$
from the $f_{SUSY}=m_{soft}^{\pm 1}$ distributions of soft terms
in the string landscape with $\mu =150$ GeV.
\label{fig:mh}}
\end{center}
\end{figure}

The distribution for the light Higgs mass is shown in Fig. \ref{fig:mh}
(taken from Ref. \cite{Baer:2022naw}).
We see for $n=1$ that the blue distribution rises to a maximum
at $m_h\sim 125$ GeV. This is where $A_t$ is large enough to yield
cancellations in the $\Sigma_u^u(\tst_{1,2})$ terms, but also lifts
$m_h$ up to $\sim 125$ GeV via maximal stop mixing\cite{Baer:2012up}.
For comparison, we
also show the orange histogram for $n=-1$ where soft terms are equally favored
at any mass scale. Here, the distribution peaks at $m_h\sim 118$ GeV
with hardly any probability at $m_h\sim 125$ GeV.

\begin{figure}[th!]
\begin{center}
\includegraphics[height=0.3\textheight]{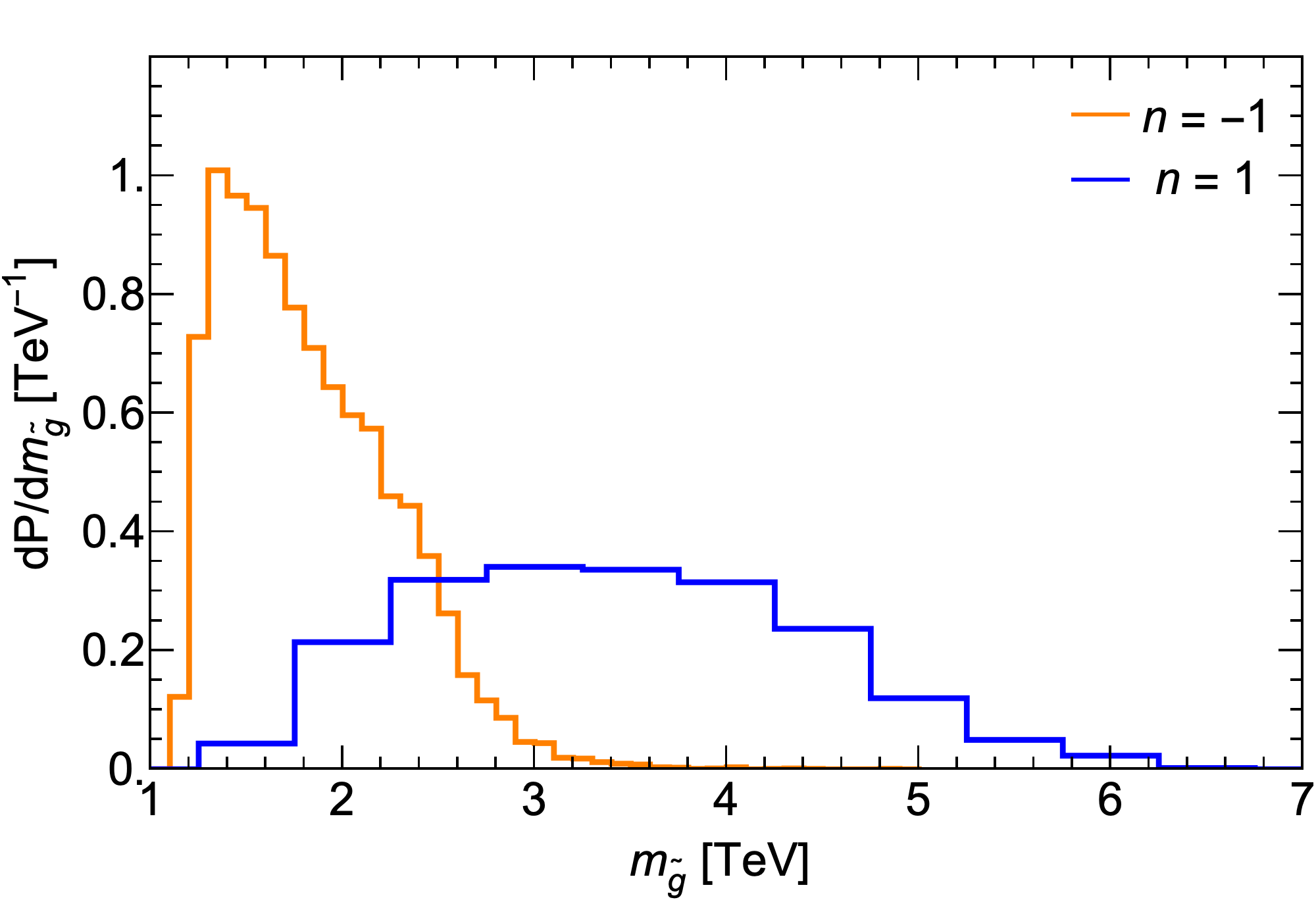}
\caption{Probability distribution for 
$m_{\tg}$ from the $f_{SUSY}=m_{soft}^{\pm 1}$ distributions of soft terms
in the string landscape with $\mu =150$ GeV.
\label{fig:mgl}}
\end{center}
\end{figure}

In Fig. \ref{fig:mgl}, we show the corresponding probability distribution
for the gluino mass. Here, for $n=1$ the curve begins around $m_{\tg}\sim 1$ TeV
and reaches a broad maximum around 3-4 TeV, while petering out beyond
$m_{\tg}\sim 6$ TeV.
The present LHC Run 2 limit from ATLAS/CMS\cite{ATLAS:2021twp,CMS:2019zmd}
is $m_{\tg}\agt 2.2$ TeV from searches within the simplified model context.
From the plot, we see that LHC is only beginning to probe the expected
range of $m_{\tg}$ values from the landscape. 

Other sparticle and Higgs mass distributions from the landscape are shown
in Ref's \cite{Baer:2022naw,Baer:2017uvn},
and they are typically beyond or even well-beyond present LHC limits.
For instance, light top-squarks are expected around
$m_{\tst_1}\sim 1-2.5$ TeV whilst first/second generation squarks and sleptons
are expected near $m_{\tq,\tell}\sim 10-30$ TeV.
From this point of view, LHC is at present seeing exactly what
the string landscape predicts.

\section{Conclusions}
\label{sec:conclude}

Theoretical arguments suggest many models which include a remnant spacetime
SUSY to populate the string landscape of $4-d$ vacua.
We assume a friendly neighborhood of the landscape populated by the MSSM
as the LE-EFT, but where the CC and also the soft SUSY breaking gaugino
masses, scalar mases and $A$-terms scan via a power-law draw to large values.
Landscape selection of soft terms then allows for a derived value of the
weak scale which must lie within the ABDS window in order for the atomic
principle to be obeyed, leading to complex nuclei and hence atoms
which are needed for complexity.

Under the landscape selection of soft SUSY breaking terms, one expects
radiative natural SUSY, or RNS, to be much more prevalent than finetuned
SUSY models such as CMSSM, G2MSSM, high scale SUSY, split SUSY or
minisplit SUSY.
This is evident because in RNS, where all contributions to the weak scale
lie within the ABDS window, there is a much larger volume of
scan space leading to $m_{weak}\in ABDS$.
Alternatively, if even one contribution to the weak scale lies outside the
ABDS window, then the remaining volume of parameter space leading to
$m_{weak}\in ABDS$ shrinks to tiny values, and is relatively less likely.
This is borne out by toy simulations of the string landscape and also allows
for a relative probability measure $P_\mu$ for different models to emerge
from the landscape.
For instance, $P_\mu (RNS)\sim 1.4$ compared to for instance
$P_\mu(HS\text{-}SUSY)\sim 6\times 10^{-4}$. Finally, we show probability
distributions of the light Higgs mass and gluinos, showing that the present
LHC is seeing what one would expect from the string landscape.
New SUSY signals especially from higgsino pair production could arise
within the next few years at LHC.
With all these beautiful results, we anticipate that
Paul will begin to work on landscape SUSY as well ;).

\section*{Acknowledgments} 

This material is based upon work supported by the U.S. Department of Energy, 
Office of Science, Office of Basic Energy Sciences Energy Frontier Research 
Centers program under Award Number DE-SC-0009956 and DE-SC-0017647.
VB gratefully acknowledges support from the William F. Vilas estate.




\bibliography{text}
\bibliographystyle{elsarticle-num}

\end{document}